\documentclass[%
 reprint,
 amsmath,amssymb,
 aps,
]{revtex4-2}
\DeclareUnicodeCharacter{2061}{}
\usepackage{graphicx}% Include figure files
\usepackage{float} %设置图片浮动位置的宏包
\usepackage{subfigure} %插入多图时用子图显示的宏包
\usepackage{dcolumn}% Align table columns on decimal point
\usepackage{bm}% bold math
\begin{document}

\preprint{APS/123-QED}

\title{Deciphering Non-Gaussianity of Diffusion Based on the Evolution of Diffusivity}% Force line breaks with \\
\author{Haolan Xu,\textsuperscript{1,3} Xu Zheng,\textsuperscript{2,*} and Xinghua Shi\textsuperscript{1,3,*}}
\affiliation{\textsuperscript{1}Laboratory of Theoretical and Computational Nanoscience, National Center for Nanoscience and Technology, Chinese Academy of Sciences, Beijing 100190, China, \textsuperscript{2}State Key Laboratory of Nonlinear Mechanics, Beijing Key Laboratory of Engineered Construction and Mechanobiology, Institute of Mechanics, Chinese Academy of Sciences, Beijing 100190, China, \textsuperscript{3}University of Chinese Academy of Sciences, No.19A Yuquan Road, Beijing 100049, China}

\email{Xu Zheng: zhengxu@lnm.imech.ac.cn\\Xinghua Shi: shixh@nanoctr.cn} 

\date{July 2023}%

\begin{abstract}
Non-Gaussianity indicates complex dynamics related to extreme events or significant outliers. However, the correlation between non-Gaussianity and the dynamics of heterogeneous environments in anomalous diffusion remains uncertain. Inspired by a recent study by Alexandre \textit{et al.} [Phys. Rev. Lett. 130, 077101 (2023)], we demonstrate that non-Gaussianity can be deciphered through the spatiotemporal evolution of heterogeneity-dependent diffusivity. Using diffusion experiments in a linear temperature field and Brownian dynamics simulations, we found that short- and long-time non-Gaussianity can be predicted based on diffusivity distribution. Non-Gaussianity variation is determined by an effective Péclet number (a ratio of the varying rate of diffusivity to the diffusivity of diffusivity), which clarifies whether the tail distribution expands or contracts. The tail is more Gaussian than exponential over long times, with exceptions significantly dependent on the diffusivity distribution. Our findings shed light on heterogeneity mapping in complex environments using non-Gaussian statistics.

\end{abstract}

\maketitle

The diffusion of microscopic particles is a key transport mechanism in various fields, including biophysics, polymer science, and composite materials. In simple fluids, diffusion is described well by the theory of Brownian motion, which predicts two fundamental features [1]: the linear variation of the mean-squared displacement (MSD) and the Gaussian displacement probability distribution (DPD). However, in complex media with heterogeneities, the Fickianity and Gaussianity assumptions are sometimes invalid [2–10]. Examples of such anomalous diffusion have been reported in living cells [3–5], polymer networks [6,7], active gels [8,9], and colloidal glasses [10].

In contrast to Gaussian behavior, non-Gaussianity indicates the presence of more complex dynamics related to extreme events or significant outliers in critical phenomena, phase transitions, and other emergent behaviors [3–14]. Notable examples include Fickian-yet-non-Gaussian diffusion (FnGD) found in actin networks and glassy materials [10–12]. The non-Gaussianity of diffusion in such complex media reflects structural or dynamical heterogeneity, and it has become a key parameter for identifying anomalous mechanisms, describing rare events, and establishing statistical inferences [13–15]. Several theoretical models based on diffusing diffusivity (DifD) have been proposed to characterize FnGD [13–19]. These models build a primary framework to show the prevalence of non-Gaussianity and illustrate the tail that is commonly assumed exponential. However, the physical meaning of non-Gaussianity in anomalous diffusion and its indication of underlying dynamics are unclear.

Because the heterogeneity of a complex environment causes the non-Gaussianity of diffusion, the central inquiry becomes understanding the correlation between the variations in non-Gaussianity and the degree of heterogeneity. The diffusivity distribution and its variation can be applied to map the structural or dynamical heterogeneity of a complex environment based on the DifD model, because they contain physical properties of the heterogeneous environment, such as viscosity, permeability, or energy barriers. The difficulty of quantifying the evolution of diffusivity, particularly in biological environments [19], makes the DifD model phenomenological and raises doubts regarding its validity. To address this problem, among many recent efforts [20–22], an inspiring approach that drew an analogy with Taylor dispersion was proposed to mathematically link non-Gaussianity and heterogeneity [22]. Nonetheless, controversies regarding variations in non-Gaussianity and tail distribution remain unresolved.

In this Letter, we used thermophoresis of nanoparticles (NPs, diameters \textit{d}=500 and 1000 nm) in a microfluidic chip [Fig. 1(a)] [23] with controlled temperature field to construct DifD of underlying probability space. We focused on clarifying the correlation between non-Gaussianity and DifD distribution\textit{ p}(D) by viewing the heterogeneous field as a spatiotemporal evolution of diffusivity. As the NPs move along the constant temperature gradient (\textit{x}-axis) with a constant thermophoretic speed $u_{T}$ [Fig. 1(b)], their diffusion, perpendicular to the temperature gradient (\textit{y}-axis), experiences DifD determined by the temperature gradient, providing better controllability and quantifiability than that reported in recent studies [20–24]. Consequently, we can predict the short-time non-Gaussianity by the type and range of the DifD distribution, and show two long-time destinations depending on the presence of external field-driven migration. We found that the temporal variation of non-Gaussianity is determined by an effective Péclet number, which identifies the competition between the varying rate of diffusivity and the diffusivity of diffusivity [13]. Unlike the majority perspective assuming an exponential tail, we use Laplace’s approximation of Bayes’ theorem to explain why a Gaussian distribution is more suitable.

\begin{figure}[H] %H为当前位置，!htb为忽略美学标准，htbp为浮动图形
\centering %图片居中
\includegraphics{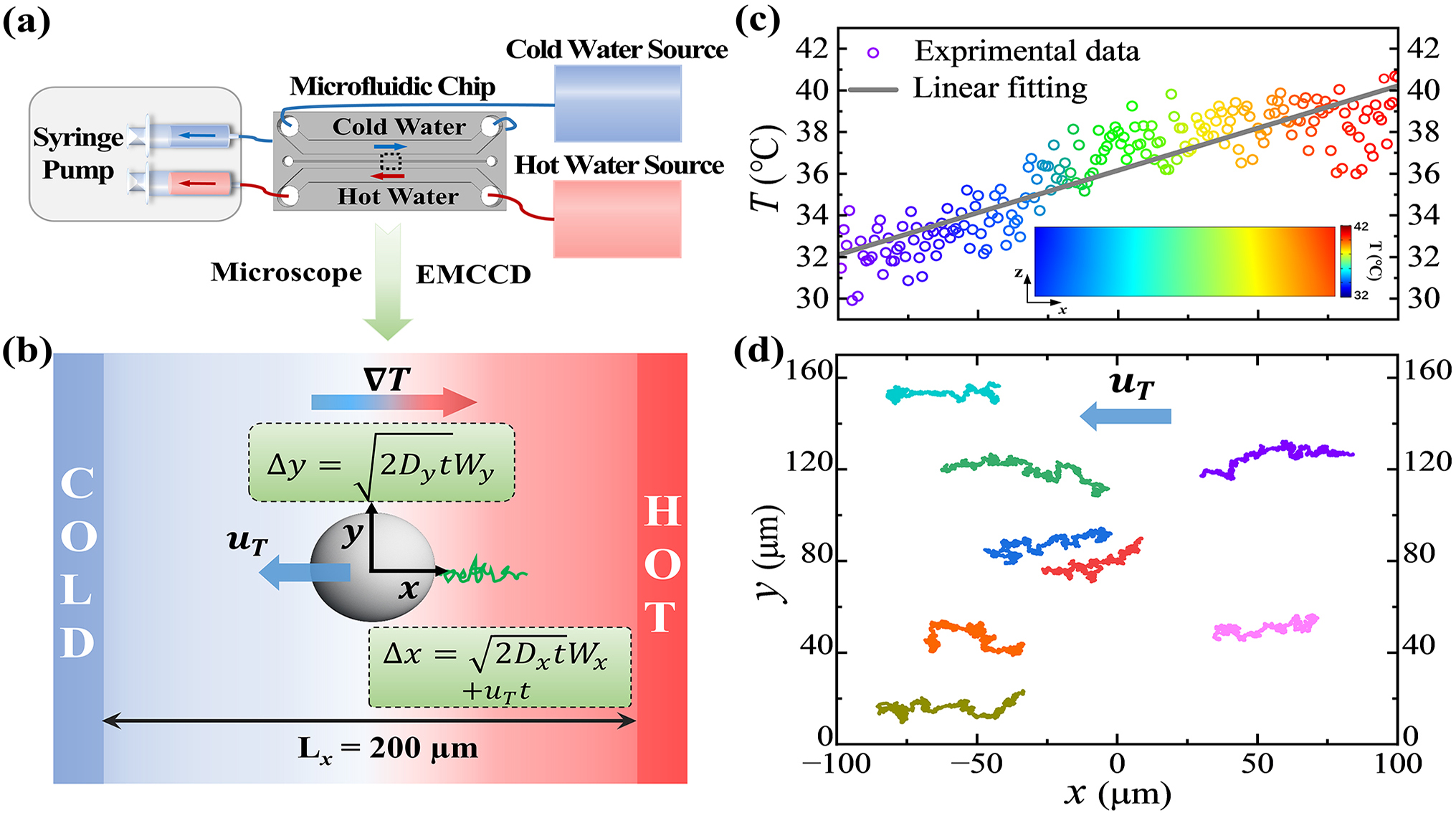} %插入图片，[]中设置图片大小，{}中是图片文件名,[width=0.48\textwidth]
\caption{(a) Experimental setup. A constant temperature gradient was created in the central microchannel of a microfluidic chip. (b) In \textit{x}-direction (spanwise of microchannel), particle moves toward the cold side along the temperature gradient, and its displacement is determined by thermophoresis and diffusion. In \textit{y}-direction (streamwise), the motion is diffusive with DifD. (c) Experimental temperature distribution in \textit{x}-direction when $\nabla\textit{T}=4.1\times10^{4}$ K/m. Inset: simulation result of the uniform temperature field [25], \textit{z}-axis is the height direction. (d) Trajectories of 1000 nm particles moving toward the cold side (left).} %A figure caption
%\label{Fig.1} %用于文内引用的标签
\end{figure}

We established stable temperature gradients of $\nabla\textit{T}=4.1\times10^{4}$ and $\nabla\textit{T}=6.6\times10^{4}$ K/m (temperature differences of 8.1 and 13.2 K) respectively in the central microchannel of the microfluidic chip by setting flow rates for cold and hot water in both side channels (Fig. 1(a)-1(c), Supplemental Material  [25]). The NPs moved toward the cold side along the x-axis with thermophoretic velocities $u_{T}$=$-0.41$ and $-0.66$ $\mu$m/s, respectively [23] [Fig. 1(d)]. At the beginning of each experiment, particles located at \textit{x}$\in$[-80, 90] $\mu$m were chosen for tracking lasting $\Delta t$=20 s.  Observation was performed at the plane of \textit{z}=15 $\mu$m to avoid wall-induced hydrodynamic drag, as the relative distance 2\textit{z/d} was large. Particle trajectories with a spatial resolution of ~80 nm were obtained from successive images captured at 20 fps [25]. The NP’s displacement in the \textit{x}-axis is $\Delta x(t)=\sqrt{2D_{x}t}W_{x}+u_{T}t+t\partial D_{x}/\partial x\thickapprox\sqrt{2D_{x}t}W_{x}+u_{T}t$, where $W_{x}$ denotes an independent stochastic process with a mean of zero and standard deviation of one, $D_{x}(T)$ is the local diffusivity depending on the temperature \textit{T(x)}, and $\partial D_{x}/\partial x$ can be neglected as it is less than 1 nm/s. In the \textit{y}-axis, the displacement follows $\Delta y(t)=\sqrt{2D_{y}t}W_{y}$, which is determined by the DifD $D_{y}(t)$ varying with NP’s thermophoretic migration along the temperature gradient in the \textit{x}-axis. 

The MSDs of 1000 nm particles under two temperature gradients are shown in Fig. 2(a), calculated using $\langle\Delta r^{2}(t)\rangle=\langle[r(t_{0}+t)-r(t_{0})]^{2}\rangle$. Here, \textit{r} represents \textit{x}

\begin{figure}[H] %H为当前位置，!htb为忽略美学标准，htbp为浮动图形
\centering %图片居中
\includegraphics{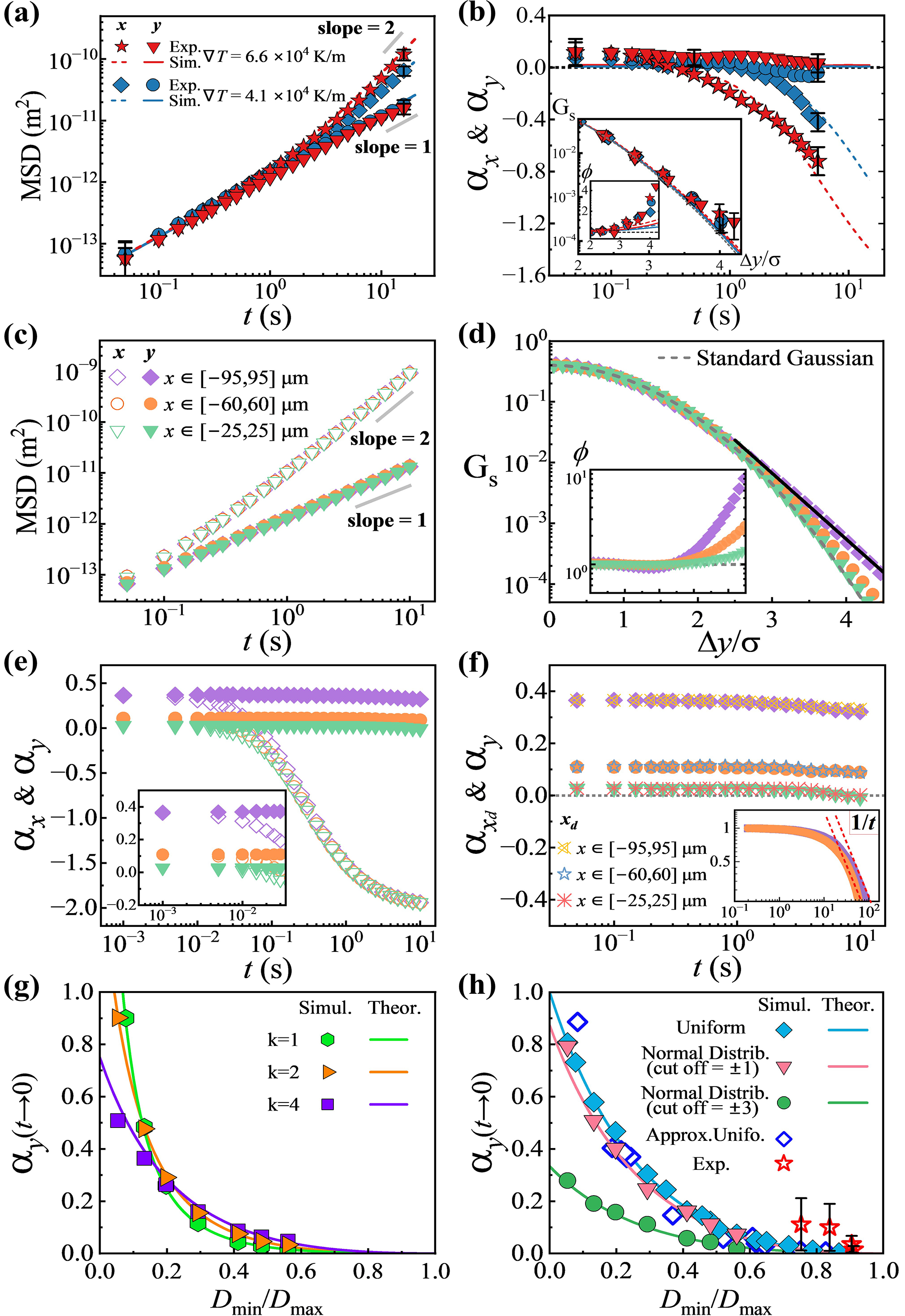} %插入图片，[]中设置图片大小，{}中是图片文件名,[width=0.48\textwidth]
\caption{Experimental (symbols) and simulation (curves) results of (a) MSD and (b) $\alpha$ for 1000 nm particles under $\nabla\textit{T}=6.6\times10^{4}$ and $\nabla\textit{T}=4.1\times10^{4}$ K/m. The inset of (b) shows normalized DPDs at $t$=0.05 s; tail ratio $\phi$ indicates the non-Gaussian tail when $\Delta y/\sigma>3$. (c)–(f) Simulation results for 1000 nm particles when $\nabla\textit{T}=3\times10^{5}$ K/m. The subscripts \textit{x}, \textit{y}, and $x_{d}$ represent \textit{x}-direction, \textit{y}-direction, and diffusive part in \textit{x}-direction, respectively. The diffusivity distributions were controlled by setting different ranges of \textit{x}. (c) MSDs. (d) DPDs at \textit{t}=0.05 s, and inset is tail ratio $\phi$. The solid line represents an exponential fit of $G_{s}\sim exp[-\vert \Delta y/(\sqrt{2D_{y}t})\vert]$. (e) Variations in $\alpha_{x}$ and $\alpha_{y}$. Inset: same initial values $\alpha_{y}$=$\alpha_{x}$ when $t\to0$. The symbols of $\alpha_{x}$ and $\alpha_{y}$ are marked based on the legend in (c). (f) Variations in $\alpha_{x_{d}}$ and $\alpha_{y}$. Inset: long-time scaling 1/\textit{t} of normalized  $\alpha_{y}$. (g)–(h) Determination of  $\alpha_{y}(t\to0)$ based on $D_{min}/D_{max}$ for (e) gamma distribution, (f) uniform distribution and truncated normal distribution.} %A figure caption
%\label{Fig.2} %用于文内引用的标签
\end{figure}

\noindent or \textit{y}, and $\langle\cdot\rangle$ denotes ensemble average. At short times, $t\thickapprox 0.1$ s, all MSDs display a linear tendency, as Brownian diffusion is dominant. When $t>2D_{x}/u_{T}^{2}\sim2$ s, the \textit{x}-MSD $\langle\Delta x^{2}(t)\rangle$ manifests a super-diffusive behavior with a slope near two, due to thermophoresis with speed $u_{T}$. By contrast, the \textit{y}-MSD $\langle\Delta y^{2}(t)\rangle$ approximates a Fickian behavior.

To investigate the non-Gaussianity of this system, normalized DPDs at \textit{t}=0.05 s were compared with the standard Gaussian distribution $G_{s}= \frac{1}{\sqrt{2\pi}}exp[-(\Delta r/\sigma)^{2}/2]$ [Fig. 2(b)], where $\sigma$ is the standard deviation. To quantify the amplification of the tailed DPD compared with the standard Gaussian distribution when $\Delta y/\sigma>3$, we show the tail ratio $\phi$  in the inset of Fig. 2(b). The maximum ratio was larger than two even though the motion in the \textit{y}-direction was purely diffusive, manifesting a non-negligible non-Gaussian tail beyond the error bars. The non-Gaussian parameter $\alpha(t)=(\langle\Delta r^{4}(t)\rangle/\langle\Delta r^{2}(t)\rangle^{2})-3$ [22] is introduced to quantitatively assess the above non-Gaussian behavior. For a Gaussian DPD, $\alpha$=0, whereas positive and negative $\alpha$ values signify fat-tail leptokurtic and platykurtic distributions, respectively. Because the uncertainty of $\alpha$ is less than $\pm 0.09$ [25], the experimental data of $\alpha_{y}$ suggest a slow decay from a positive $\alpha_{y}\thickapprox0.12$ when $t<0.1$ s to $\alpha_{y}\thickapprox0$ when $t\thickapprox1$ s. When $t>1$ s, $\alpha_{y}$ is still considered to be zero based on the error bars, which is further confirmed by numerical simulations. By contrast, $\alpha_{x}$ starts from a small positive value, similar to $\alpha_{y}$, but decays more rapidly to a negative value when $t>1$ s. Greater temperature gradient ($\nabla\textit{T}=6.6\times10^{4}$ K/m) causes larger non-Gaussian parameter $\alpha_{y}$ and fatter tail at short times. Experiments using 500 nm NPs show similar MSD and DPD results [25]. 

To tackle the issue of limited long-time statistics in the experiments, we used Brownian dynamics simulations [25] to complement the long-time evolution of non-Gaussianity. By assigning the same thermophoretic speed and uniform diffusivity distribution along the \textit{x}-axis as the experiment, the simulation results [dashed and solid curves in Fig. 2(a)–2(b)] closely match the experimental data. The long-time $\alpha_{x}(t)$ from the simulation collapses onto the extension of the experimental data when $t\sim5$ s. Next, primarily based on the simulation results, we explore the effect of the spatiotemporal variation of DifD on non-Gaussianity. The initial distribution of $D_{r}$ and its varying rate $\partial\langle D_{r}\rangle/\partial t\sim(\partial\langle D_{r}\rangle/\partial T)\nabla Tu_{T} $ are controlled [25], and the diffusivity of diffusivity, defined as $\partial\langle D_{r}^{2}\rangle/\partial t$, and the varying range of $D_{r}$ are monitored. 

We first test the non-Gaussianity at $\nabla\textit{T}=3\times10^{5}$ K/m. After subjecting the particles to thermophoresis with $=-3.0$ $\mu$m/s for $\Delta t$=15 s, the diffusivity distributions when ensemble statistics was conducted were regulated by setting the ranges of \textit{x} positions to $[-95, 95]$ $\mu$m, $[-60, 60]$ $\mu$m, and $[-25, 25]$ $\mu$m, respectively, from the initial positions \textit{x}$\in$$[-50, 95]$ $\mu$m, $[-15, 60]$ $\mu$m, and [20, 25] $\mu$m (SM [25]). A broader \textit{x} range indicates a wider diffusivity distribution. In addition to the typical MSDs [Fig. 2(c)] and DPDs [Fig. 2(d)], similar to the experimental tendency, the variations in the non-Gaussian parameters in both directions are shown in Fig. 2(e)–2(f). The short-time  $\alpha_{y}(t)$ (solid symbols) starts from approximate constant values of 0.37, 0.11, and 0.03, following a gradual decay at $t>2D_{y}/u_{T}^{2}\sim0.1$ s. Interestingly, the non-Gaussianity shown by green symbols is similar to the experimental result with $\nabla\textit{T}=6.6\times10^{4}$ K/m [Fig. 2(b)], because they have similar diffusivity ranges though different $\nabla\textit{T}$. For a wider diffusivity distribution, the value of $\alpha_{y}(t)$ is larger and the tail is broader. The value of $\alpha_{x_{d}}(t)$[Fig. 2(f)], based on $\Delta x_{d}=\Delta x-u_{T}t$, is always almost the same as $\alpha_{y}(t)$. Interestingly, $\alpha_{x}(t)$ [empty symbols in Fig. 2(e)] rapidly turns negative and reaches $-2 $ at $t\sim10$ s, despite shares the same beginning as $\alpha_{y}(t)$. 

The above results demonstrate the significant influence of diffusivity distribution on non-Gaussianity and fat-tailed DPD. Recalling a recent study suggesting an analogy of DifD motion with Taylor dispersion [22], we find that the asymptotic features of non-Gaussianity can be accurately predicted only if the distribution of $D_{r}$ is known. The mathematical derivation by Alexandre \textit{et al.} [22] gives the fourth cumulant as:  $\langle\Delta r^{4}\rangle-3\langle\Delta r^{2}\rangle^{2}=12\langle[\int_{0}^{t} ds(D_{r}-\langle D_{r}\rangle)]^{2}\rangle$. When $t\to0$, the non-Gaussian parameter can be approximately calculated by dividing the variance $\textbf{\textit{Var}}(D_{r})$ over the square of expectation $\langle D_{r}\rangle^{2}$ [22,25]:
\begin{equation}
\alpha_{r}(t\to0)=3(\langle D_{r}^{2}\rangle-\langle D_{r}\rangle^{2})/\langle D_{r}\rangle^{2}=3\textbf{\textit{Var}}(D_{r})/\langle D_{r}\rangle^{2}
\end{equation}

Eq. (1) correlates $\alpha_{r}(t\to0)$ with the diffusivity distribution \textit{p}($D_{r}$) in a probability space, which explains the same initial values $\alpha_{y}$=$\alpha_{x}$=$\alpha_{x_{d}}$ in Fig. 2(e)-2(f) because they share the same diffusivity. This relation could be applied to various systems with structural or dynamical heterogeneities by mapping the heterogeneity onto DifD. We modify distribution \textit{p}($D_{r}$) from uniform distribution, which is suitable for our experiments, using uniform sampling, to other commonly-used distributions. Surprisingly, we find that $\alpha_{y}(t\to0)$ can be predicted solely based on the range ratio $\beta(\Delta t)=D_{min}/D_{max}$. We take a gamma distribution $\textit{p}(D_{y})=D_{y}^{k-1}exp(-D_{y})/\Gamma(k)$ as an example, which is truncated at $D_{y}\in[D_{min}, D_{max}]$. By mapping it to a standard gamma distribution $\textit{p}(z)=z^{k-1}exp(-z)/\Gamma(k)$ truncated at $z\in[0, N]$: $(D_{y}-D_{min})/(D_{max}-D_{min})=z/N$, we obtain $\textbf{\textit{Var}}(D_{y})=\textbf{\textit{Var}}(z)(D_{max}-D_{min})^{2}/N^{2}$ and $\langle D_{y}\rangle=\langle z\rangle(D_{max}-D_{min})/N+D_{min}$. For large $N$ such as $N$=10, $\textbf{\textit{Var}}(z)\thickapprox k$ and $\langle z\rangle\thickapprox k$, $\alpha_{y}(t\to0)$ of the gamma distribution is:
\begin{equation}
\alpha_{y}(t\to0)=3k[\frac{1-\beta}{k(1-\beta)+N\beta}]^{2}
\end{equation}

To verify Eq. (2), we assigned gamma distribution to $D_{r}$ in simulations and maintained $\nabla\textit{T}$ and $u_{T}$ unchanged. The simulation results [Fig. 2(g), $\Delta t$=0.5 s] for different shape parameters, $k$=1, 2, and 4, were in good agreement with the prediction curve from Eq. (2). The result of gamma distributions can be extended to other exponential or power-law distributions and can be widely used to evaluate non-Gaussianity in complex scenarios. Besides, we derive the expressions of $\alpha_{r}(t\to0)$ for uniform and normal distributions as $\alpha_{y}(t\to0)=(\frac{1-\beta}{1+\beta})^{2}$ and $\alpha_{y}(t\to0)=\frac{3}{N^{2}}(\frac{1-\beta}{1+\beta})^{2}$, respectively, which perfectly match the simulation data in Fig. 2(h). By mapping to a standard normal distribution truncated at $z\in[N, N]$, $\alpha_{y}(t\to0)=\frac{3}{N^{2}}(\frac{1-\beta}{1+\beta})^{2}$ is valid when $N\ge3$. The expression of $\alpha_{y}(t\to0)$ when $N<3$ is given in SM [25], which predicts a larger non-Gaussianity owing to a stronger dispersion of $D_{y}$ [pink curve in Fig. 2(h) for $N$=1]. Note that $\alpha_{y}(t\to0)$ in our simulation only varies slightly with the duration $\Delta t$ as the change of $\beta(\Delta t)$ is tiny during slow evolution of diffusivity distribution, which has been manifested by the short-time plateau of $\alpha_{y}(t)$ in Fig. 2(e)-2(f). The data (empty blue diamonds) are close to the theoretical curve of uniform distribution for $\Delta t$=10 s when the diffusivity distribution is still approximately uniform. The experimental data acquired from the uniform samples were located near our prediction curve in Fig. 2(h) based on uniform distribution, suggesting the good predictive power of our approach up to $\Delta t$=20 s. Furthermore, our results illustrate that non-Gaussianity in the same system can vary significantly depending on the sampling range of $p(D_{r})$. In particular, ergodic sampling with a minimal value of $\beta$ can exhibit a much larger non-Gaussianity than truncated sampling with a larger $\beta$, as illustrated in Fig. 2(g–h).

We then consider the long-time limit of non-Gaussianity in Fig. 2(e–f). Although calculating the temporal evolution of $p(D_{r})$ is complicated for various heterogeneities, previous work [22] has predicted that $\alpha_{r}(t)$ should decay with $1/t$ at long times. We indeed observed a $1/\textit{t}$ decay to zero of $\alpha_{y}$ when $t\sim$100 s [inset of Fig. 2(f)] in our simulation, when boundary confinement was absent. In realistic cases with confinement [21,22], $1/\textit{t}$ decay was observed later when $t\sim$1000 s. This long-time scaling $1/\textit{t}$ is assumed to be an intrinsic feature owing to diffusivity dispersion, which is independent on boundary confinement. 

Distinct from the positive non-Gaussianity, the most evident feature of $\alpha_{x}(t)$ is the asymptotic value $\alpha_{x}(t\to \infty)=-2$ owing to the thermophoresis. Substituting $\Delta x(t)=\sqrt{2D_{x}t}W_{x}+u_{T}t$ into $\alpha_{x}(t)=(\langle\Delta x^{4}\rangle-3\langle\Delta x^{2}\rangle^{2})/\langle\Delta x^{2}\rangle^{2}$, one obtains $\alpha_{x}(t)=(4\langle D_{x}^{2}\rangle/u_{T}^{4}t^{2}-12\langle D_{x}\rangle^{2}/u_{T}^{4}t^{2}-2)/(4\langle D_{x}^{2}\rangle/u_{T}^{4}t^{2}+4\langle D_{x}\rangle^{2}/u_{T}^{4}t^{2}+1)$. The long-time limit is $\alpha_{x}(t\to \infty)=-2$ if $\langle D_{x}^{2}\rangle\ll u_{T}^{4}t^{2}$. An intriguing inference is that the asymptotic value $\alpha_{x}(t\to \infty)=-2$ will appear for any external field-driven migration with a large speed $u_{T}^{4}t^{2}\gg\langle D_{x}^{2}\rangle$.

Unlike the experiment which reported a rapid increase in non-Gaussianity at short times [21], our experimental and simulation results show a constantly decreasing $\alpha_{y}(t)$. Next, we discuss the tendency of $\alpha_{y}(t)$ by the derivative $\frac{\partial\alpha_{y}(t)}{\partial t}\sim\frac{1}{\langle D_{y}\rangle^{2}}(\frac{\partial\langle D_{y}^{2}\rangle}{\partial t}-2\alpha_{y}\langle D_{y}\rangle\frac{\partial\langle D_{y}\rangle}{\partial t})$, which shows a competition between $\frac{\partial\langle D_{y}^{2}\rangle}{\partial t}$ and  $\frac{\partial\langle D_{y}\rangle}{\partial t}$, referred to as the diffusivity of diffusivity and the varying rate of diffusivity respectively. This competition defines an effective Péclet number $Pe=(\langle D_{y}\rangle\frac{\partial\langle D_{y}\rangle}{\partial t})/\frac{\partial\langle D_{y}^{2}\rangle}{\partial t}$, which helps assess whether $\alpha_{r}(t)$ will increase or decrease with time. Here, $\langle D_{y}\rangle$, $\frac{\partial\langle D_{y}\rangle}{\partial t}$, and  $\frac{\partial\langle D_{y}^{2}\rangle}{\partial t}$ can be seem as equivalent length, velocity, and diffusivity,  respectively,  in a

\begin{figure}[H] %H为当前位置，!htb为忽略美学标准，htbp为浮动图形
\centering %图片居中
\includegraphics{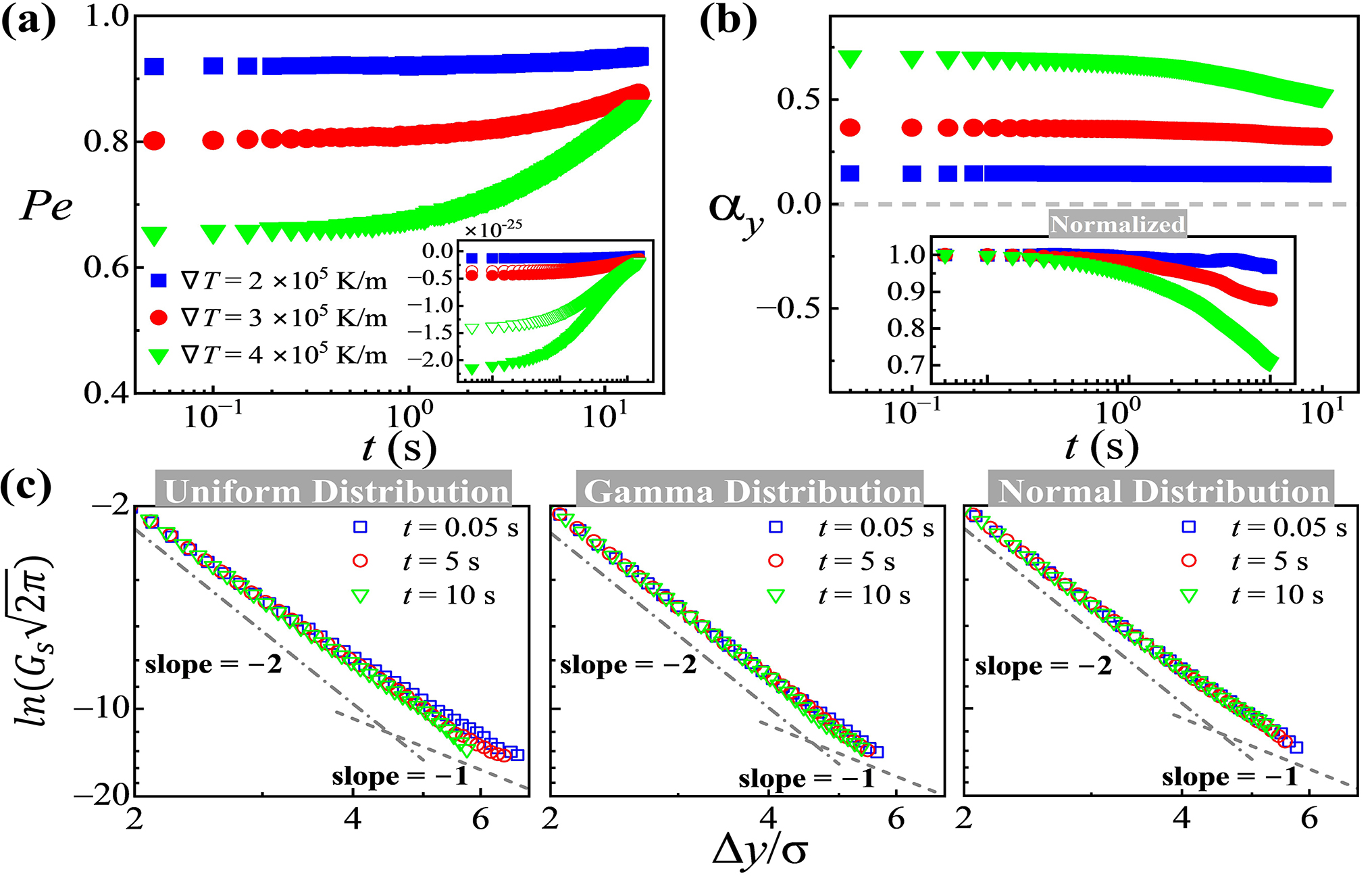} %插入图片，[]中设置图片大小，{}中是图片文件名,[width=0.48\textwidth]
\caption{(a) Temporal variation of \textit{Pe} under different temperature gradients, Inset: $\frac{\partial\langle D_{y}^{2}\rangle}{\partial t}$ (solid symbols) and $\frac{\partial\langle D_{y}\rangle^{2}}{\partial t}$ (open symbols). (b) $\alpha_{y}(t)$ and corresponding curves normalized by the maximum values shown in inset. (c) $ln(G_{s}\sqrt{2\pi})$ \textit{vs} $\Delta y/\sigma$ for different distributions $p(D_{y})$ at $t$=0.05–10 s. The slope of $-2$ means $G_s$ is Gaussian, whereas $-1$ means $G_s$ is exponential.} %A figure caption
%\label{Fig.3} %用于文内引用的标签
\end{figure}

\noindent space of $D_{y}$, in analogy to traditional $Pe$ number in Taylor dispersion. As the varying rate can be approximated as $\frac{\partial\langle D_{y}\rangle}{\partial t}\sim(\partial\langle D_{y}\rangle/\partial T)\nabla Tu_{T}\sim\nabla T^{2}\sim u_{T}^{2}$, in Fig. 3(a–b) we adjust the temperature gradient $\nabla T$ to investigate its influence on the variations of $Pe$ and $\alpha_{y}$. Both the diffusivity of diffusivity $\frac{\partial\langle D_{y}^{2}\rangle}{\partial t}$ (solid symbols) and the varying rate $\frac{\partial\langle D_{y}\rangle^{2}}{\partial t}$ (open symbols) are negative [inset of Fig. 3(a)], whereas the domination of the diffusivity of diffusivity, i.e., $|\frac{\partial\langle D_{y}^{2}\rangle}{\partial t}|>|\frac{\partial\langle D_{y}\rangle^{2}}{\partial t}|$, results in $Pe<1$ and $\frac{\partial\alpha_{y}(t)}{\partial t}<0$. Larger $\nabla T$, which increases the degree of heterogeneity, causes a larger  $\alpha_{y}(t)$ and a smaller $Pe$ number as the ratio $|\frac{\partial\langle D_{y}\rangle^{2}}{\partial t}|/|\frac{\partial\langle D_{y}^{2}\rangle}{\partial t}|$ decreases with increasing $\nabla T$. The competition between $\frac{\partial\langle D_{y}^{2}\rangle}{\partial t}$  and $\frac{\partial\langle D_{y}\rangle}{\partial t}$ depicts the following two-stage variation: at short times, the leading contribution $\frac{\partial\langle D_{y}^{2}\rangle}{\partial t}$ is dominated by diffusion, resulting in slow variations of both $Pe$ and $\alpha_{y}(t)$ and a smaller $Pe$ number; and at intermediate times when thermophoretic motion is dominant over diffusion, $\frac{\partial\langle D_{y}\rangle}{\partial t}\sim u_{T}$ becomes significant, causing a fast increase of $Pe$ number and a fast decay of $\alpha_{y}(t)$.

Interestingly, above analysis predicts a positive $\frac{\partial\alpha_{y}(t)}{\partial t}$ at short times if $\frac{\partial\langle D_{y}^{2}\rangle}{\partial t}$ and $\frac{\partial\langle D_{y}\rangle^{2}}{\partial t}$ turn positive while $|\frac{\partial\langle D_{y}^{2}\rangle}{\partial t}|>|\frac{\partial\langle D_{y}\rangle^{2}}{\partial t}|$ is maintained. We obtained positive $\frac{\partial\alpha_{y}(t)}{\partial t}$ in the simulation by changing the NP from thermophobic to thermophilic [25]. Nonetheless, the increase in short-time non-Gaussianity in our simulation was much weaker than the rapid dynamics reported by Pastore \textit{et al}. [21]. We speculate that the difference was due to a sudden dispersion of $\frac{\partial\langle D_{r}^{2}\rangle}{\partial t}$ produced by the optical-illumination speckle [21]. The different non-Gaussian behaviors between the present system with slow DifD and the systems with rapid dynamics, such as glassy materials [10], can serve to detect specific short-time mechanisms.

After clarifying the correlation between non-Gaussianity and distribution $p(D)$, we further discuss how $p(D)$ determines fat-tailed DPD, which is the most prominent feature of FnGD. Whether the tail of the DPD is exponential remains controversial [7,12,22,26], although a good exponential fit is observed in Fig. 2(d). Mathematically, the DPD $G_{s}(\Delta y,t)=\int p(D_{y)}p(\Delta y|D_{y})dD_{y}$ following Bayes' theorem can be rewritten as $G_{s}(\Delta y,t)=\int_{D_{min}}^{D_{max}}\frac{p(D_{y}(x))}{\sqrt{4\pi D_{y}(x)t}}exp⁡(-\frac{\Delta y^{2}}{4D_{y}(x)t})dD_{y}$. $D_{y}(T)$ is determined by the temperature field. This integral can be approximated as $G_{s}(\Delta y,t)\thickapprox Cexp⁡[-\frac{\Delta y^{2}}{4D_{y}(x^{*})t}]$ based on the Laplace approximation for large $\Delta y$, where prefactor $C$ depends on $p(D_{y})$, and $x^{*}$ is the position of maximum $D_{y}(x^{*})$. This approximation could help clarify the controversy regarding the tail as the term $exp⁡[-\frac{\Delta y^{2}}{4D_{y}(x^{*})t}]$ suggests a more Gaussian tail for large $\Delta y$, which contradicts many existing results of exponential tails [6,7,11,21,26]. As shown in the double-logarithmic plot of $ln(G_{s}\sqrt{2\pi}) vs. \Delta y/\sigma$ in Fig. 3(c), the slopes of gamma and normal distributions are approximately $-2$(Gaussian) for the tail $\Delta y/\sigma>3$. The function type of $p(D_{y})$ can influence the result of the integral by prefactor $C$ in the Laplace approximation. For the uniform distribution of $p(D_{y})$, the slope of the tail deviates from $-2$ when $\Delta y/\sigma>3$ and becomes $-1$ near $\Delta y/\sigma=6$. This is consistent with the prediction of an exponential tail if $\partial ln(p(D_{y}))/\partial D_{y}$ is constant, as reported by Wang \textit{et al. }[12], using the steepest descent analysis for the integral $G_{s}\sim\int exp[ln(p(D_{y}))-\frac{\Delta y^{2}}{4D_{y}t}]/\sqrt{D_{y}}\cdot dD_{y}$. Although the Gaussian tail for the gamma distribution is distinct from the prediction in [12,13], it is in accordance with Alexandre \textit{et al.} [22], where a generic Gaussian tail was proposed. Additionally, a slight contraction of the tail turning more Gaussian with time, is observed for uniform and gamma distributions as the Gaussian term $exp⁡[-\frac{\Delta y^{2}}{4D_{y}(x^{*})t}]$ becomes more significant for large $\Delta y$. The transition from a short-time exponential-like tail to a long-time Gaussian-like tail could clarify the controversy that similar systems may display contradictory tail shapes. As an exception, diffusion in strongly confined media [7], where statistical data cannot reach a sufficiently large $\Delta y$, typically manifests an exponential tail. 

\textit{Conclusion.}—We investigated non-Gaussian diffusion in a thermophoretic system with controlled DifD and correlated non-Gaussianity with the spatiotemporal evolution of diffusivity distribution. Short-time non-Gaussianity was predicted based on the range ratio $\beta$ of diffusivity distribution. We demonstrated the predictive power using uniform, normal, and gamma distributions. The variation in non-Gaussianity at intermediate times was determined by the effective $Pe$ number, which characterizes the competition between the varying rate of diffusivity and the diffusivity of diffusivity. This explains why the tail of the DPD usually contracts, unless it is influenced by sudden dynamics that enhance the diffusivity of diffusivity. The long-time decay of non-Gaussianity followed a $1/t$ scaling in a purely diffusive process, whereas in the presence of particle migration, the non-Gaussianity eventually reached $-2$. Furthermore, using Laplace’s approximation of Bayes’ theorem, we depicted that the tail was more Gaussian than exponential. We showed the transition from a short-time exponential-like tail to a long-time Gaussian-like tail to clarify the controversies in existing experiments. Our results shed light on heterogeneity mapping in complex environments based on non-Gaussian statistics, echoing the idea that Bayesian deep-learning deciphers the physics encoded in diffusion data [27].
\begin{acknowledgments}
This work was supported by the Strategic Priority Research Program of the Chinese Academy of Sciences (XDB36000000), the National Key R\&D Program of China (2022YFA1203200, 2022YFF0503504), the Natural Science Foundation of Beijing (2222085, 1202023, 2194092), the National Natural Science Foundation of China (12072350, 11972351, 12102456, 11672079, 12072082, 12125202). The authors thank Hefei Advanced Computing Center for computational resources.

\end{acknowledgments}

\nocite{*}

% 参考文献列表

\end{document}